\pdfoutput=1  
\documentclass[10pt,conference]{IEEEtran}
\IEEEoverridecommandlockouts

\usepackage{cite}
\usepackage{amsmath,amssymb,amsfonts}
\usepackage{graphicx}
\usepackage{textcomp}
\usepackage{xcolor}
\usepackage{enumitem}
\usepackage{bbm}
\usepackage{dsfont}
\usepackage{siunitx}
\usepackage{float}
\usepackage{xurl}    
\usepackage{hyperref}

\begin{document}

\title{Towards Deploying Optimistic Quantum Fourier Transforms: An Architecture-Algorithm Co-Design Study}

\author{
\IEEEauthorblockN{Pedro L. S. Lopes}
\IEEEauthorblockA{QuEra Computing Inc. \\
1380 Soldiers Field Road, Boston, MA - USA \\
plopes@quera.com}
}

\maketitle

\begin{abstract}
We present an architecture--algorithm co-design study of the Optimistic Quantum Fourier Transform (OQFT) under a surface-code fault-tolerant execution model for reconfigurable neutral-atom hardware. Analyzing the OQFT structure---particularly its reliance on phase-gradient resources and small-scale blocks---highlights architectural requirements for resource mobility and parallel execution. Guided by that, we introduce a \emph{hot-zone} architecture that decouples data storage from processing and dynamically routes mobile resource packages (magic-state factories, bridge qubits, and phase-gradient registers) to stationary data regions. To expose dominant costs, we route rotation insertions via catalytic phase-gradient addition and heuristically micro-schedule ripple-carry adders to patch-level moves. Under this model, leading Gidney~\cite{Gidney2018halvingcostof} and Cuccaro~\cite{cuccaro2004} adders are found to exhibit similar space--time volume but require different levels of parallelism. At the algorithm level, the five-layer OQFT shows a tunable parallelism/latency trade-off: two hot zones match serial-QFT latency, four hot zones roughly halve runtime, and additional hot zones asymptotically approach constant-time execution at substantial resource cost. Across $256$--$2048$-bit instances, the requirements for half-time performance converge to about $500$ additional logical ancillae and a peak parallelism of $128$ logical qubits. We also identify broader algorithm--architecture bottlenecks, including endianness mismatches between phase-gradient and data registers, which we address via cyclic phase-gradient swaps and alternating QFT reflections. Scoped to surface codes and cultivation-only magic-state factories, our analysis identifies reaction-limited operation and parallelism demand as primary drivers of resource estimation and establishes a generalizable foundation for primitive-based architectural studies.
\end{abstract}


\section{Introduction}

A simple heuristic to estimate the duration of a fault-tolerant quantum algorithm takes the form

\begin{align}
    \text{T}_{\text{algo}}=\max{\left( \text{T}_{\text{react}},\text{T}_{\text{magic}} \right)}. \label{eq:time_heuristics}
\end{align}
Here, $\text{T}_{\text{react}}$ denotes the reaction-limited algorithm time— the integral time to run the protocol with a direct concatenation of operations in units of the reaction time $t_r$, the minimum time between co-dependent adaptive measurements—while $\text{T}_{\text{magic}}$ denotes the time required to generate all the magic states needed by the algorithm. The former sets the fastest possible execution time given the limits of a system's classical control stack, while the latter reflects the requirement that the algorithm cannot complete until all non-Clifford resources have been produced. The role of these time scales has been extensively studied, both jointly and independently, particularly because high-quality magic-state generation has historically remained a major bottleneck for fault-tolerant quantum computation~\cite{fowler2013timeoptimalquantumcomputation, gidney2020quantumblocklookaheadadders}. However, Eq.~\ref{eq:time_heuristics} omits several important factors, including the relative balance between Clifford and non-Clifford gates, the complexity of information routing, and the cost of parallelism. Recent advances in magic-state generation~\cite{gidney2024magicstatecultivationgrowing,chen2025efficientmagicstatecultivation,sahay2025foldtransversalsurfacecodecultivation,menon2025magictricyclesefficientmagic} and studies of architectural connectivity~\cite{litinski2022activevolumearchitectureefficient,Harry_AFT}, both at the logical and physical levels, are beginning to highlight the importance of a more fine-grained analysis of the interaction between algorithms and quantum hardware.

Careful accounting of Clifford and non-Clifford gate counts, scheduling constraints, and information routing within realistic system layouts has already led to substantial improvements in resource estimates for large-scale quantum protocols such as Shor's algorithm~\cite{harry_architecture,gidney2025factor2048bitrsa,cain2026shors10000,babbush2026ECC,Litinski2019magicstate,Gouzien_2023,litinski2022activevolumearchitectureefficient,webster2026pinnaclearchitecturereducingcost}. These studies also demonstrate how architectural constraints can feed back into algorithm design, introducing examples that include transversal measurement-based gadgets for long-distance CNOT fanout and lattice-surgery-aware adder constructions. Such careful co-design remains difficult to automate, as it typically requires detailed considerations specific to a given hardware platform, compilation strategy, and error-correction framework. In this work, we focus on one concrete instance of that challenge: a procedural co-evaluation of a relevant routine and an explicit architectural model, aimed at identifying the hardware conditions under which the algorithmic trade-off is practically beneficial.

As our example we consider the optimistic quantum Fourier transform (OQFT), currently the principal example of an optimistic quantum circuit~\cite{kahanamokumeyer2025logdepthinplacequantumfourier}. This class of circuits introduces controlled ``algorithmic error''~\cite{gidney2019approximateencodedpermutationspiecewise} in order to reduce circuit resources, accepting bounded average error over the Hilbert space under the Frobenius metric. In the case of the OQFT, this approximation reduces the depth of the standard linear-depth QFT to logarithmic depth, at the cost of additional Clifford and non-Clifford operations. The primary proposed application of the OQFT is in reducing the depth of fast quantum multipliers~\cite{kahanamokumeyer2024fastquantumintegermultiplication}. The OQFT offers opportunities for increased parallelism and simplified gadgetization, advantages whose practical value may depend strongly on the availability of independent control over many degrees of freedom and the supply of resource states (e.g., magic factories and catalytic states). The features above make the OQFT a particularly good example for our study, though the methodology we adopt in this work can be deployed to more general algorithms. The central question we address in this work is therefore: \emph{under which architectural requirements can the OQFT trade-off be made beneficial?}

We concretely consider a reconfigurable neutral-atom platform implementing the surface code with transversal fault tolerance~\cite{Harry_AFT}. While architectures based on high-rate qLDPC codes are gaining significant, due, attention~\cite{yang2026spacetimeefficienthardwarecompatiblecomplexquantum,webster2026pinnaclearchitecturereducingcost,xu2024fastparallelizablelogicalcomputation,xu2025batchedhighratelogicaloperations,cain2025correlated}, we conduct our analysis using the aforementioned surface-code-based architecture as it offers a concrete and operationally well-understood baseline for co-design exploration, which has also been shown to support large-scale algorithm execution on neutral-atom platforms~\cite{harry_architecture}. At the same time, this architecture exhibits operational speed constraints that motivate exploration of strategies for faster algorithmic primitives. Moreover, previous architectural studies merit to be revisited in light of recent advances in magic-state cultivation~\cite{gidney2024magicstatecultivationgrowing,chen2025efficientmagicstatecultivation,sahay2025foldtransversalsurfacecodecultivation}. In this work, we leverage this setting to develop and exercise a procedural co-design approach, uncovering how architectural constraints shape algorithmic structure and performance. In doing so, we characterize the interplay between parallelism, ancilla provisioning, magic-state supply, subroutine selection, and pipelining allowing up to a $2\times$ reduction in OQFT execution time relative to standard implementations while remaining within realistic, though intensive, hardware bounds.

In what follows, we review the structure of the OQFT and present a scaffolded analysis of how to deploy it efficiently. We identify algorithmic properties that are particularly relevant for neutral-atom platforms, highlight opportunities for efficient routing, and explore strategies for reducing operations and improving pipelining. Our analysis connects high-level resource estimates with more detailed microscheduled models. Along the way, we highlight several practical techniques for routing and scheduling, including: (1) using ancillas as factories in Gidney-style adders~\cite{Gidney2018halvingcostof}; (2) saving time steps in logical-AND constructions via swaps; and (3) pipelining sequential $\text{QFT}$ blocks using cyclic swaps of phase-gradient (PG) adders and alternation with $\text{QFT}^\dagger$ blocks to maintain resource alignment and minimize atom movement. A final observation is that the practical performance gap between leading adder constructions may be smaller than commonly reported~\cite{gidney2020quantumblocklookaheadadders}. When other factors are comparable, the degree of parallelism required to sustain performance may instead become the decisive factor in determining algorithmic viability.
\begin{figure*}[t]
    \centering    \includegraphics[width=\textwidth]{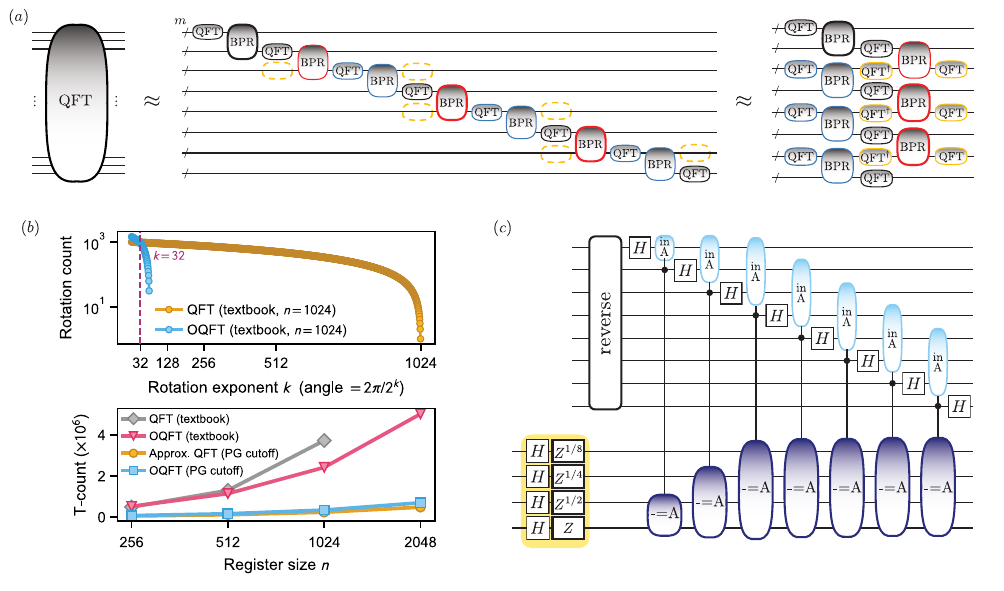}
    \caption{\textbf{Logic and resource structure of Optimistic Quantum Fourier Transforms. }(a)~An $n$-qubit QFT is truncated and decomposed into $m\!\sim\!\log n$ sub-blocks and block-phased rotations (BPR) $e^{iXY/2^{2m}}$, where $X$ and $Y$ are the register contents of the blocks involved. Approximately commuting the blue sub-circuits past the red ones---at the cost of extra yellow log-sized QFT blocks---yields the log-depth OQFT~\cite{kahanamokumeyer2025logdepthinplacequantumfourier}. (b)~Rotation spectra (top) and T-counts (bottom) for different QFT and OQFT implementations. The OQFT naturally truncates small rotations relative to the textbook QFT, but introduces ${\sim}44\%$ additional rotations. Catalytic additions into a truncated phase-gradient (PG) state, as shown in~(c), eliminate exponentially small rotations from both implementations. With PG, the approximated textbook QFT has slightly lower T-counts than the OQFT; moreover, a single PG state re-serializes the OQFT, raising the question of how to realize its log-depth promise in practice. OQFT block sizes are 32 bits, and angles are truncated at $2\pi/2^{32}$ throughout. (c)~PG-based QFT circuit for an $8$-bit register with rotations truncated at $Z^{1/8}$. $A$ indicates the adder input values in superposition on corresponding register. This technique make controlled additions the main computational bottleneck of implementing the protocol.}
    \label{fig:OQFT_overview}
\end{figure*}

\section{OQFT overview} \label{sec:OQFT_overview}

Figure~\ref{fig:OQFT_overview}(a) illustrates the sequential steps in defining the OQFT circuit. Starting from a regular QFT applied to a register, we break the register into blocks of size $m=\log n/\epsilon'$, where $n$ is the register size and $\epsilon'$ is the target error bound uniformly averaged across the Hilbert space~\cite{kahanamokumeyer2025logdepthinplacequantumfourier}. The result is a set of intra-block log-sized QFTs together with ``block-phased rotations'' (BPR)---$Z$-phase rotations of the form $e^{i 2\pi XY/2^{2m}}$, where $X$ and $Y$ are the $m$-bit binary representations of the control and target block contents~\cite{kahanamokumeyer2025logdepthinplacequantumfourier}. Approximate commutations are then performed by inserting additional log-sized QFT blocks that act as phase estimators, yielding independence between blocks and thus enhanced parallelism at smaller depths. These phase estimations fail, however, at the edges of the represented value range, because $m$-bit binary arithmetic wraps around modularly. The resulting failures are severe but confined to a small region of the Hilbert space. Targeting register sizes between $n=256$ and $2048$, with error bounds $\epsilon'<10^{-4}$, fixed to target percent-level chance of failure in a factoring protocol, we settle for $m=32$ throughout this work, rounding $m$ up towards the closest relevant power of 2 to ensure divisibility in $n/m$ and avoid ``ragged tail'' small blocks.

Besides depth reduction, this strategy has other desirable consequences. Notably, the smaller registers of the log-sized QFT and BPR blocks naturally eliminate a large fraction of the irrelevant phase rotations present in the textbook QFT---that is, rotations with phases $\pi/2^k$ for large $k$, cf.\ Fig.~\ref{fig:OQFT_overview}(b), top panel; obtained via Qualtran~\cite{qualtran}. Despite an increase in the number of larger rotations---namely the blue points below $k=32$, arising from the extra inserted small QFTs (yellow in Fig.~\ref{fig:OQFT_overview}(a))---substantial savings in T-counts relative to the textbook QFT emerge as the register size grows, as shown in Fig.~\ref{fig:OQFT_overview}(b), bottom panel (grey vs.\ pink). Ultimately, these savings stem from the fact that the number of rotations in the textbook QFT scales as $\mathcal{O}(n^2)$, whereas the OQFT rotation count scales as $\mathcal{O}(nm)\sim \mathcal{O}(n\log n)$.

Naturally, this is not a fair comparison, as we are allowing the OQFT to incur approximation errors but not the textbook QFT. Approximating the textbook QFT by removing angles below an error target is a well-documented procedure~\cite{coppersmith2002approximatefouriertransformuseful}. Because the OQFT incurs roughly $44\%$ more rotations due to the extra small QFT blocks, the approximated textbook QFT consistently outperforms the OQFT in T-counts. In what follows, we consider a particular variant in which explicit rotation synthesis is replaced by catalytic additions into a phase-gradient (PG) state~\cite{jones2013thesis,gidney2017approximateqftblog,qualtran2025approximateqft}. Specifically, the $k$-qubit phase-gradient state $|\nabla_k\rangle \equiv 2^{-k/2}\sum_{x=0}^{2^k-1} e^{2\pi i x/2^k}|x\rangle$ satisfies the catalytic identity $\text{ADD}(a)\,|\nabla_k\rangle = e^{-2\pi i\, a/2^k}\,|\nabla_k\rangle\,$, verifiable by sending $x\to x+a$ in $|\nabla_k\rangle$ and reordering the sum. So, adding an integer $a$ (quantum or classical) into the register applies a $Z$-rotation of angle $2\pi a/2^k$ while leaving the state intact. Each rotation that would otherwise require $\mathcal{O}(\log 1/\epsilon)$ T gates for direct synthesis is replaced by an addition into the PG register, whose Toffoli cost is independent of the rotation angle; BPRs are themselves made of a series of controlled phase gradients, so the same technique extends to BPR rotations. The circuit for an 8-bit QFT with rotations truncated at $Z^{1/8}$ is shown in Fig.~\ref{fig:OQFT_overview}(c) (using subtractions rather than additions; equivalent up to register negation and a unit increment, with negligible magic cost). The blue and yellow curves in the lower panel of Fig.~\ref{fig:OQFT_overview}(b) confirm that the PG implementation achieves considerably lower T-counts than the textbook implementations of both the regular QFT and the OQFT. However, although the PG state helps eliminate remaining irrelevant BPR rotations, the additional rotations inherent to the OQFT result in a T-count systematically above that of the approximated linear QFT.

It is worth considering these gate counts in the context of the main proposed use case of the OQFT: fast multipliers for factoring~\cite{kahanamokumeyer2024fastquantumintegermultiplication}. These multipliers perform arithmetic over quantum rotation phases, reusing techniques from efficient classical multipliers such as Karatsuba and Toom-Cook. Reference~\cite{kahanamokumeyer2024fastquantumintegermultiplication} estimates roughly $0.9 \times 10^6$ Toffolis and $0.1 \times 10^6$ rotations per invocation of the fast phase-based multiplier (for $n=2048$) using QFTs, which appear at the boundaries of the multiplier circuit and limit its depth,  assumed to use PG implementations. We opt for a 4T per Toffoli decomposition~\cite{Jones4T_toffoli} and, under a per-rotation synthesis error budget of $\epsilon=\epsilon'/n = 10^{-5}/n$---yielding $85$--$90\%$ factoring success for $n=1024$ and $2048$ respectively---Qualtran's synthesis model gives about $40$ T gates per rotation~\cite{qualtran}. The full standard Shor pipeline for factoring requires $2n$ controlled modular multiplications for the phase-estimation resolution; accounting for the controlled operations overhead (a factor of two in non-Clifford cost), the total T-count reaches the order of $10^{9}$--$10^{10}$. Inverting this count gives a per-T error target of roughly $10^{-10}$--$10^{-11}$.

Following literature, we assume physical error rates of $p_\text{phys}\sim 10^{-3}$~\cite{harry_architecture}, which make the per-T error target above challenging for magic-state cultivation alone. State-of-the-art fold-transversal cultivation~\cite{sahay2025foldtransversalsurfacecodecultivation} at fault distance $f=5$ reaches per-T errors of roughly $10^{-10}$, and would suffice for executing Shor's factoring algorithm on registers up to $n\approx 512$ with ``fast multipliers''. It becomes, however, only marginally sufficient for that at $n=1024$, and surely insufficient at $n=2048$. In principle, a single round of $8T\!\to\!\text{CCZ}$ distillation quadratically suppresses the input error and restores comfortable margins at all register sizes, but it also introduces additional factory footprint and latency~\cite{Litinski2019magicstate}. This extra complexity would furthermore compound with the inevitable practical tensions the use of PG also introduces. A PG register, despite being catalytic, is a single resource that must be shared across the system. If only one copy is deployed in an OQFT setting, the protocol is effectively re-serialized---with the added disadvantage of carrying extra rotations. A performant architecture must therefore be able to replicate such resources while managing them efficiently. 

We therefore scope our architectural analysis to cultivation-only factories, accepting that the results apply directly only to registers of up to $\sim\!512$ bits in the factoring with ``T-heavy fast-multipliers'' context. Results for $1024$ and $2048$ are reported as well, but would only matter for validation experiments on the OQFT. This scope is nevertheless rich enough to expose the key architectural trade-offs we wish to study, allowing us to consider perspectives on register alignment, efficient pipelining, and a thorough revision of adder performance comparisons as shown below.

\begin{figure*}[t]
    \centering    \includegraphics[width=\textwidth]{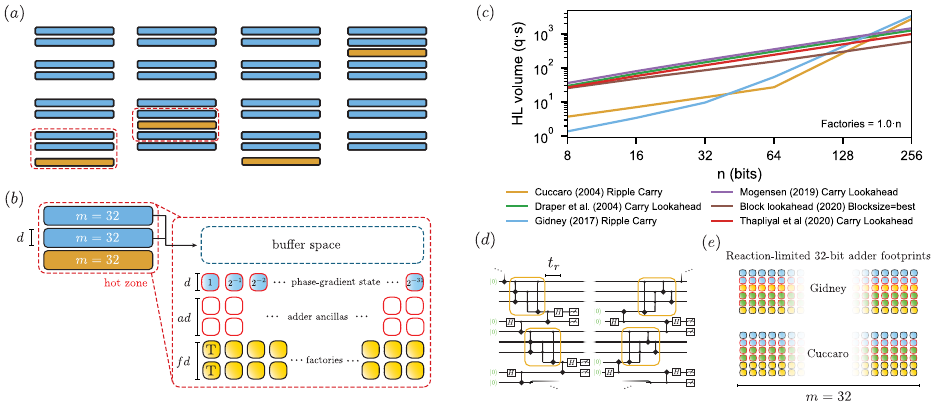}
    \caption{\textbf{Coarse-grained architecture overview and reaction-limited adder resource analysis review. }(a)~Macro-scale architecture: the system register is organized into $m\!=\!32$ data blocks (blue), paired vertically. Here four pairs are displayed side by side. Surface-code logical qubits are assumed throughout. Resources (dark yellow) form a mobile gadget that, when docked to data blocks above or below, defines a hot zone as shown in~(b), which can drive processing. (b)~Anatomy of a hot zone: rows of ancillae (red-edged squares), magic-state factories (yellow, black-edged; fold-transversal cultivation), and phase-gradient registers. The buffer space is occupied by data qubits as needed. (c)~Revisited circuit volume for a comprehensive set of adders under Gidney’s high-level (HL) resource model~\cite{gidney2020quantumblocklookaheadadders}; with neutral-atom time scales and modern cultivation factors, the crossover to lookahead adders shifts to $\sim128$-bit registers, though ripple-carry remains favorable. (d)~Bridge-qubit interleaving in logical-AND-driven MAJ and UMA adder blocks: Bell pairs enable semi-parallel execution with up to a $t_r$ shift between consecutive operations. (e)~Footprints for no-stall, reaction-limited operation of Gidney-style (logical-AND with measurement-based uncomputation) and Cuccaro-style ($7T$ Toffoli synthesis) adders. Green red-edged squares are bridge-qubit ancillae; yellow red-edged ancillae in Gidney's layout correspond to logical-AND ancillae. Cuccaro's higher magic demand requires an additional factory row to avoid stalling, ultimately equalizing the register footprints of the two protocols at the sizes of interest.}
    \label{fig:macro_and_adderHL}
\end{figure*}

\section{Macro level architecture}\label{sec:architecture_macro}

To evaluate compilation strategies for the OQFT, we must fix a hardware architecture and a fault-tolerance framework. We consider a neutral-atom system with surface codes and transversal operations under the transversal fault-tolerance (TFT) paradigm~\cite{Harry_AFT,harry_architecture}. Each logical gate is followed by a single syndrome-extraction (SE) round and decoded via correlated decoding across entangled code patches~\cite{cain2025correlated,Harry_AFT}. The surface-code framework provides a well-established basis for layout and microschedule design, and its patch geometry aligns naturally with the spatial requirements of magic-state cultivation~\cite{sahay2025foldtransversalsurfacecodecultivation}. A key consequence of TFT is that the QEC cycle time coincides with the reaction time, $t_\text{QEC}=t_r$---a relation that will matter when we compile the most resource-intensive sub-routines. We pick representative physical and error correction parameters from typically accepted values in the literature; Table~\ref{tab:params} collects them and the corresponding references of origin.

\begin{table}[t]
\centering
\caption{System and error-correction parameters.}
\label{tab:params}
\begin{tabular}{l c c}
\hline\hline
Parameter & Value & Ref.\\
\hline
Physical error rate $p_\text{phys}$ & $10^{-3}$ &  \cite{harry_architecture}\\
Code distance $d$ & 15 & ---\\
Site spacing $\ell$ & $12\;\mu$m & \cite{harry_architecture}\\
Acceleration $a$ & $5500\;\text{m/s}^2$ & \cite{harry_architecture}\\
Physical gate time $t_\text{gate}$ & $1\;\mu$s & \cite{harry_architecture}\\
Measurement time $t_\text{meas}$ & $500\;\mu$s & \cite{harry_architecture}\\
Decoding time $t_\text{dec}$ & $500\;\mu$s & \cite{harry_architecture}\\
Reaction time $t_r = t_\text{QEC}$ & $1\;\text{ms}$ & ---\\
Cultivation fault distance $f$ & 5 & \cite{sahay2025foldtransversalsurfacecodecultivation}\\
Avg.\ cultivation attempts & ${\sim}8$ & \cite{sahay2025foldtransversalsurfacecodecultivation}\\
Successful cultivation period $t_\text{cult}$ & $10\;\text{ms}$ & ---\\
\hline\hline
\end{tabular}
\end{table}

\emph{Code distance and error budget:}
Under TFT, the per-gate logical error rate is modestly elevated relative to pure memory due to the more complex correlated decoding problem. Using the numbers from Ref.~\cite{harry_architecture}, the transversal CNOT, the least performant Clifford gate in this setting, incurs a per-operation error of $p_{L,\text{CNOT}} \approx 7\times 10^{-9}$ at $d=15$ and $p_\text{phys}=10^{-3}$. On the non-Clifford side, we take fold-transversal cultivation at $f=5$ yielding ${\sim}10^{-10}$ error per T state. While embedding the OQFT in a full Shor pipeline would require slightly larger distances and magic state distillation, Fig.~\ref{fig:OQFT_overview}(b) suggests that our choice suffices to validate OQFTs with PG cutoffs even as large as $2048$ bits. With improvements, e.g. in decoders, these numbers could also be pushed down to $d=13$ or below, at least for validation purposes. Finally, the reaction time $t_r = t_\text{meas} + t_\text{dec} = 1\;\text{ms}$ sets the fundamental clock of the system, the rate at which non-Clifford operations can proceed. We pick a high-level estimate of the cultivation period as $t_\text{cult}=10\;\text{ms}$, reflecting the ${\sim}8$ average post-selection attempts required at $f=5$; at ${\sim}10$ reaction ticks per T state, factory throughput will be a central scheduling constraint in what follows.

\emph{Shuttling and processing dynamics:} Atom transport times are derived from the ballistic formula $t_\text{move}=2\sqrt{L/a}$~\cite{harry_architecture}, where $L$ is the physical distance; for a single code-patch span ($L = d\,\ell = 180\;\mu$m) this gives $t_\text{move}\approx 360\;\mu\text{s}$, comfortably within one reaction tick. Conversely, inverting the formula for $t_{\text{move}}=1\;\text{ms}$ yields a reach of $\sim7$ patch spans per reaction tick, beyond which a qubit in ballistic transit must pause for syndrome extraction. Our engine shifts all move times by a constant value due to qubit pick and drop times, currently estimated at $400\; \mu\text{s}$~\cite{Manetsch_2025,fast_transport,Pause_2024,Bluvstein_2022_coherent}. This shift, however, does not affect our simulations strongly, getting absorbed by the reaction-limited operation of our adders described below.

\emph{Hot/data zoning}: At this stage, we are ready to state our architecture of choice. We adopt a top-down approach: first a macro-level layout, then detailed microschedules for most relevant sub-routines. At the highest level, we partition the system into functionalized zones: stationary \emph{data zones} and mobile \emph{hot zones}, each dedicated to a specific role. Figure~\ref{fig:macro_and_adderHL}(a,b) illustrates the layout. Data qubits are grouped in blocks of $m=32$---a power of two close to the optimal OQFT block size for the register sizes we target---and arranged in vertical pairs, mirroring the structure of the BPR and small-QFT operations. A hot zone sits between a pair of data blocks and comprises of all resources needed for the efficient operation of the algorithm of choice, in our case the magic-state factories, adder ancillas, and a PG register; it can operate on the data registers immediately above or below. The design principle is that \emph{resources and magic move to data, not the reverse}: the hot zone hops vertically from pair to pair while data registers remain stationary. With appropriate routing and pipelining, this enables for shorter individual shuttling distances, as well as fewer routing conflicts. Naturally, further dynamics happen inside a hot zone once it is parked close to its data targets.

We layout the hot zones and adjust the number of ancillas and factories as needed specifically for our main process of interest: addition into phase gradients. A high-level (HL) modelling alternative, previously considered in the literature e.g. in Ref.~\cite{gidney2020quantumblocklookaheadadders}, is to treat magic-state factories as abstract and freely available resources that can serve any part of a circuit. While this successfully captures HL details, this can overcount effective throughput and neglects the complexity of routing magic and resources for efficient computing, such as bridge qubits~\cite{fowler2013timeoptimalquantumcomputation}, to where they are needed---an issue that is compounded in the OQFT, where the PG states must also be shared through data blocks. Conversely, one may argue that our approach model is too tightly coupled to a particular sub-routine to generalize across an entire application, and may still waste resources in part of a computation. Finding the right balance between specificity and generality is an open challenge in quantum systems integration; in this work, we pursue designs that are as aligned as possible with the sub-routine of interest.

Aligning the data as a long horizontal strip minimises the hot-zone travel distance per hop, but field-of-view limitations in optical systems may force the strip to be split into side-by-side columns. This incurs a limited number of inter-column moves whose distance scales logarithmically with the register size. As we determine below, vertical hops between the data and factory rows traverse about ${\sim}6$ patch spans (${\sim}890\;\mu\text{s}$, likewise within one QEC cycle). When internal dynamics lead to moves that take longer than a QEC cycle, logical patches must stop mid movement for a syndrome-extraction round before computation resumes; we account for that by effectively doubling the cost of such movement times in units of reaction ticks. In our simulations, we only consider the situation where the system is aligned on a long strip; since a more square form-factor can be obtained with a couple of divisions of that strip for any sizes of interest, the number of log-long horizontal moves, however, is always order $\mathcal{O}(1)$ for our protocol. Those can be batched in groups of about 7 logical qubits and, overall would lead to irrelevant extra costs. Likewise, our model purposefully does not account for space for qubit measurement; due to flexibility on form factors for that, we don't expect that to considerably change conclusions. Technology alternatives might also allow for in-situ measurement schemes that could negate the need of such considerations~\cite{Anand_2024}.

\begin{figure*}[t]
    \centering    \includegraphics[width=\textwidth]{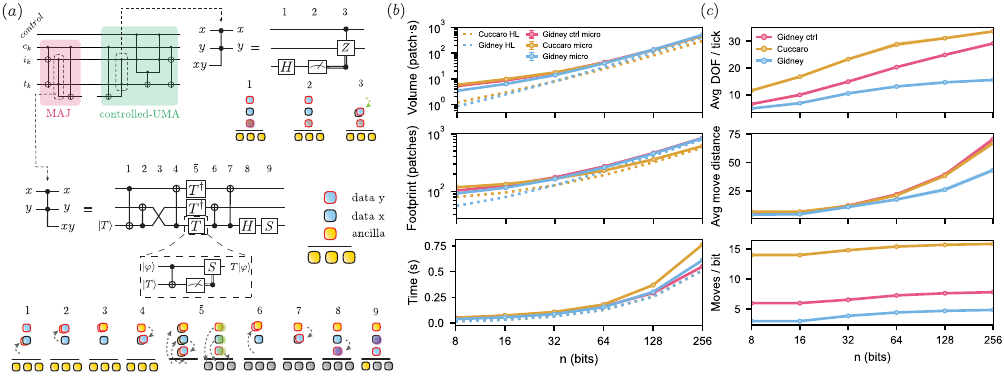}
    \caption{\textbf{Fine-grained adder scheduling and simulated performance. }(a)~Microscheduled routing for one iterative block of Gidney's controlled adder. Face and edge colors identify qubit roles. Yellow patches with black edges are T~factories, grey patches are reset after measurement, and purple shading marks single-qubit gates. Green operations are classically controlled and count at half depth in our cost model. The logical-AND ancilla is assumed to start with a cultivated T~state. The ${\sim}10$ steps of a logical-AND operation roughly match the T-state re-cultivation time, facilitating factory pipelining. Relative to Ref.~\cite{Gidney2018halvingcostof}, we eliminate one CNOT via a near-free SWAP during MAJ. (b)~Performance comparison of our microscheduled Gidney (controlled and uncontrolled) and Cuccaro adders against the high-level (HL) requirements of Fig.~\ref{fig:macro_and_adderHL}(c), adapted to our cost model. With resources sized for stall-free, reaction-limited operation, all microscheduled adder variants achieve near equal circuit volume; discrepancies with the HL analysis matter more at smaller registers. (c)~Average degrees of freedom (DOF) moved per tick, move distances, and move counts a bit experiences break the performance degeneracy. Cuccaro's deeper MAJ and UMA blocks keep more qubits simultaneously in flight on average, resulting in higher parallelism requirements.}
    \label{fig:micro_data}
\end{figure*}

\section{Micro level architecture}

The implementation of an OQFT reduces to sequential controlled additions onto a PG register. Performance is a function of the addition strategy, where many options arise. With advances in magic-state production, it is worth revisiting conclusions from earlier adder analyses. In Figure~\ref{fig:macro_and_adderHL} we reproduce the HL cost model of Ref.~\cite{gidney2020quantumblocklookaheadadders} with production rates and parameters updated to our neutral-atom model with magic-state cultivation. Several quantum adder circuits are compared, falling broadly into two categories: carry-lookahead and ripple-carry adders. The cost metric is the protocol volume in (logical) qubit-seconds including factory counts, assumed to scale with the addition register size. In this volume cost model, adding factories increases both the qubit footprint and the magic production rate in a way that the two effects partially cancel in the volume metric: doubling the factory count roughly doubles the footprint but also halves the T-limited execution time.

The conclusions of Ref.~\cite{gidney2020quantumblocklookaheadadders} still hold: ripple-carry adders outperform lookahead adders, as magic remains a scarce resource. However, the crossover point in register size has shifted down by an order of magnitude, while the volume gap between the two families remains about one order of magnitude. Ultimately, lookahead adders struggle due to their extra magic resource requirements. Although the high-level nature of this cost model may overestimate the performance gap, these observations gives us sufficient confidence to focus on ripple adders, which, although considered in the context of neutral-atom transversal architectures, have not yet been analyzed with state-of-the-art cultivation-style factories~\cite{harry_architecture,sahay2025foldtransversalsurfacecodecultivation}.

Key limitations of the cost model of Ref.~\cite{gidney2020quantumblocklookaheadadders} include the absence of routing constraints, the omission of ancilla registers needed to improve the reaction depth of adders, and an indirect accounting of factory footprint. We revise the model following the design philosophy of our specialized hot zone, including in the footprint the exact resources needed for optimized adder operation from $2$ up to $32$ bits. We consider Gidney's~\cite{Gidney2018halvingcostof} and Cuccaro's~\cite{cuccaro2004} ripple-carry adders, which are technically the same algorithm differing only in how they implement the AND operations. By Gidney's adder we mean Toffoli implementation via an ancilla and $4$ T gates with measurement-based uncompute, while by Cuccaro's we mean the direct Toffoli implementation with $7$ T gates and no ancilla---specifically the synthesis shown in Fig.~3b of Ref.~\cite{dutta2025}~\footnote{Technically, Cuccaro's adder does require one carry-in ancilla. This is a negligible cost that can be absorbed in our T-factory ancillas without generating meaningful stalling. Truly ancilla-independent linear adder constructions also exist and may be directly considered if desired~\cite{Takahashi:2005ygp}}. We incorporate Bell pairs for time-optimal computation~\cite{fowler2013timeoptimalquantumcomputation,harry_architecture}, cf.\ Fig.~\ref{fig:macro_and_adderHL}(d). This allows compressing the interleaving of majority (MAJ) and unmajority-and-add (UMA) blocks with a single reaction tick delay, at the cost of two extra qubits---bridge qubits---per register bit. Our model does not explicitly account for the frame-update costs of operating bridge qubits, nor for other reaction-limited gadgets such as auto-T/CCZ~\cite{gidney2019flexiblelayoutsurfacecode}. Our volume, is simply the total footprint multiplied by the total estimated time of the algorithmic primitive of interest.

Several observations are worth highlighting. First, our model suggests that stall-free operation of these adders at register sizes of interest requires \emph{the same footprint} (Fig.~\ref{fig:macro_and_adderHL}(e)). While Gidney's adder requires an extra ancilla per bit (blue squares, red-lined), Cuccaro's is about twice as T-gate intensive, requiring more factory patches (yellow squares, black-lined). These factors offset each other, yielding identical hot-zone footprints.

At the microschedule level, Gidney's logical-AND (LAND) operations offer several optimization opportunities. Highlights of our proposed microschedules, summarized in Fig.~\ref{fig:micro_data}(a), include (i)~the removal of a CNOT gate from the original LAND via a swap operation, which on neutral atoms incurs only an extra pick-and-drop of logical qubits, reducing the step count by one (step~3); (ii)~the observation that cultivation provides one factory per patch, so LAND ancillae can be initialized directly with cultivated T states, saving one teleportation step; and (iii)~the coincidence that the ${\sim}8$ steps of the LAND operation approximately match the T-gate regeneration time, facilitating factory pipelining.

The circuit depths of the MAJ and UMA blocks are also important. While bridge qubits allow displacing these blocks by a single reaction depth regardless of adder variant, the depth of the semi-parallelized blocks determines how many are simultaneously in flight, and thus how many bridge qubits must be active. Deeper blocks imply more concurrent initializations before any one resolves. The total depths suggested by our model are $15.5$ reaction ticks for Gidney (MAJ: $11$, UMA: $4.5$) and $24$ for Cuccaro (MAJ: $12$, UMA: $12$), computed assuming classically-controlled operations count at half depth (since they are not always implemented). We find that bridge-qubit recycling is not beneficial at the sizes considered: deep MAJ blocks require so many bridge qubits that by the time a pair becomes free, transporting it ahead is not worthwhile. Overall, stall-free operation of adders from $2$ to $32$ bits (and slightly beyond) requires two complete $32$-patch rows of bridge qubits, plus one row of magic factories and one row of ancillae for Gidney, or two rows of factories for Cuccaro---yielding the matched footprint noted above. The smaller register adders could naturally perform well with fewer resources, but the convenience of our layout given the algorithm of interest remains.

Figure~\ref{fig:micro_data}(b) summarizes the comparison across volume, footprint, and time. Both protocols operate at the reaction limit and, with the matched footprints noted above, achieve comparable---though not identical---volumes. Cuccaro's deeper blocks cost ${\sim}20\%$ more time, but its smaller data-register footprint ($2n$ vs.\ $3n$) increasingly compensates: the Cuccaro-to-Gidney volume ratio drops from ${\sim}1.8\times$ at $n=8$ to ${\sim}1.2\times$ at $n=32$, crossing unity near $n=64$--$128$. Our adapted HL model (Fig.~\ref{fig:macro_and_adderHL}(c)) already hinted at this convergence but predicted the crossover at $n=32$ and underestimated absolute volumes by $2$--$5\times$ at small sizes, where routing overhead and factory idle time dominate. In practice, register sizes beyond $32$ bits are rarely encountered in monolithic form: windowing~\cite{gidney2019windowedquantumarithmetic} and the OQFT itself~\cite{kahanamokumeyer2025logdepthinplacequantumfourier} decompose typical arithmetic circuits into $5$--$32$~bit sections~\cite{gidney2025factor2048bitrsa,Gidney2021howtofactorbit,litinski2023compute256bitellipticcurve,Gouzien_2023}, placing the regime of interest squarely where the two adders are hardest to distinguish by volume alone. Because LAND ancillae are cleaned after uncompute and reused, the same conclusion extends to the controlled adder (Fig.~\ref{fig:micro_data}(b), top left), which is our actual primitive of interest.

Where the two protocols \emph{do} differ is in their parallelism demands. Like any depth-reducing technique, bridge qubits demand a capacity to control and move more concurrent degrees of freedom (DOF). We quantify this via the number of DOF per tick of the computation clock (one QEC cycle). Importantly, the number of simultaneously active bridges scales with the depth of the blocks being stitched. To approximately avoid stalling on both models, we provision a cap of 40 DOF per tick, and monitor the average number of active logical qubits. As Fig.~\ref{fig:micro_data}(c) shows, DOF analysis breaks the volume degeneracy decisively: Gidney's shallower MAJ and UMA blocks require considerably fewer average DOF per tick than Cuccaro's. From a hardware implementation perspective, this extra requirement is key due to the footprint required by light deflection technology, both in physical real state and optical path. We should also note that while we evenly provisioned our simulations with a cap of 40 DOF per tick on both adders, Gidney's version has a smaller peak DOF requirement for no-stalling operations, and actually caps at only $32$ DOF.

Move distances become relevant only at larger sizes and, owing to the ripple-carry structure and randomized nature of cultivation-based factories, are dominated by factory-to-compute transfers; these average about $12$ patch spans for $n=32$ bits (though under only $5$ spans for $n=8$) indicating that magic injection could frequently require mid-move syndrome extraction. That may be avoided as long as idling errors are subleading, so one might not need to refresh qubits at each cycle; determining whether this is the case would require detailed noisy simulations we leave for future studies. The number of moves per bit follows similar trends. The parallelism requirements for the controlled adder interpolate between those of Gidney's and Cuccaro's, settling at around $15$ DOF per tick at $n=32$, with comparable average move distances. Weighing all factors, we select the controlled Gidney adder for integration into the OQFT protocol, though we stress that this choice is not clear-cut.
\section{Integrated performance}

\begin{figure}[t]
    \centering    \includegraphics[width=\columnwidth]{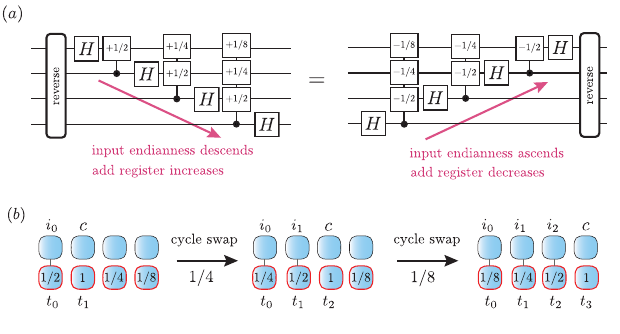}
    \caption{\textbf{Mesoscale routing analysis. }(a)~Two near-equivalent realizations of a QFT: reversing the addition order requires only a register negation and a unit increment. In both cases the most-significant target bit shifts as the protocol progresses. (b)~Cyclic swapping of the phase-gradient register (red edges) keeps addition inputs and targets aligned. At our distance and time scales, swaps can be batched in groups of seven within a single QEC cycle. Note that PG additions within a QFT block have a fixed most-significant bit $i_{n-1}=0$, not displayed. Most reverse operations can be commuted through the OQFT resulting in a single group of rearrangements in the beginning or end of protocol.}
    \label{fig:endianess}
\end{figure}

We now compile the full OQFT protocol and evaluate its integrated performance. A first issue concerns the alignment of endianness between the PG state and the data registers, most clearly visible in the small-QFT blocks. As shown in Fig.~\ref{fig:endianess}, the data qubit serving as the most-significant-bit target changes at each controlled PG rotation, yet the PG state is spatially ordered with its most significant bit (the $Z^{1/2}$ rotation) at a fixed position. Consequently, as the addition register size grows sequentially, the input for a given rotation drifts farther from its PG target, creating long-distance routing requirements. Our solution is a progressive cyclic swap of the PG state as the calculation evolves (Fig.~\ref{fig:endianess}(b)). This keeps input and target aligned while minimising long-distance moves. The cycling requires short shifts of the PG register, which can be batched in groups of $6$--$7$ logical patches and completed in at most five or six iterations for the largest adders---a negligible overhead. 

PG cycling has a further consequence when the full algorithm is considered. While BPR operations commute (they require no interleaved Hadamards), the sequential small-block QFTs impose an ordering on additions from $2$- to $32$-bit registers. An endianness conflict arises because cycling pushes the large-angle rotations away from the most-significant data bit when additions always start at $2$ bits. The solution is to reflect alternate QFT sub-blocks (Fig.~\ref{fig:endianess}(a), right panel), replacing additions with subtractions into the PG register. This costs only a register negation and a unit increment, negligible in magic. Alternating the QFT direction as small sub-blocks progress through the OQFT keeps addition inputs and targets aligned while maintaining qubit-shuttling requirements under control. We note that this endianess mismatch problem is generic and architecture-independent. Our cyclic permutation and pipeline routing-scheme however are specific to the architecture of choice and highlight the power of reconfigurable technology. 

With these alignment issues resolved, our architecture allows us to ``gadgetize'' the OQFT implementation, exploiting its high parallelism while managing ancillae and magic resources. We integrate the full protocol into our model, parameterized by the number of hot zones (hz). A single hot zone yields a serialized baseline; adding more hot zones up to the architectural maximum ($\text{hz}=\text{max}$, set by the number of block pairs) unlocks the OQFT's parallelism. For example, a $256$-bit OQFT has four pairs of $32$-bit blocks, admitting at most $\text{hz}=4$; this maximum doubles with the register size.

\begin{figure}[t]
    \centering    \includegraphics[width=\columnwidth]{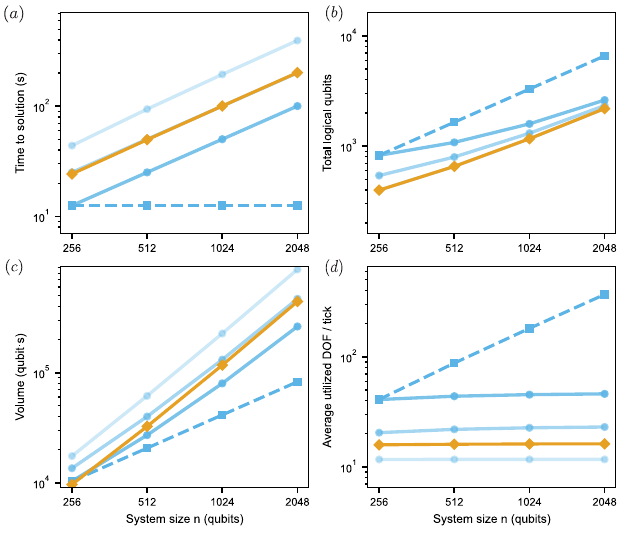}
    \caption{\textbf{Integrated OQFT performance as a function of register size for varying hot-zone counts. } Yellow diamonds mark the regular (serial) QFT baseline. (a)~Time to solution. Saturating the hot-zone count ($\text{hz}=\text{max}$) yields constant-time execution at the cost of a ${\sim}6\times$ increase in logical qubits; at fixed counts, $\text{hz}=2$ suffices to match the regular QFT, while $\text{hz}=4$ halves the execution time. (b)~Qubit footprint. (c)~Combined protocol volume as product of footprint-time. (d)~Average concurrent degrees of freedom (DOF). At $\text{hz}=4$, the overhead is roughly $500$ logical ancillae and average $46$ concurrent logical operations.}
    \label{fig:OQFT_finals}
\end{figure}

When the hot-zone count is not saturated, the OQFT time to solution scales polynomially with system size, as does the regular QFT (Fig.~\ref{fig:OQFT_finals}(a)). The fixed slopes indicate a constant multiplicative speed-up that nearly halves the time each time the hot-zone count doubles. For sizes ranging from $256$ to $2048$ bits, $\text{hz}=2$ suffices to break even with the regular QFT, compensating the extra magic cost through a somewhat significant increase in parallelism requirements. Saturating the hot-zone count yields constant-time scaling. That is a natural consequence of the fixed five-layer OQFT structure and fixed 32-bit sub registers. Unfortunately, achieving this performance also incurs a steep cost in ancillae and factories (Fig.~\ref{fig:OQFT_finals}(b)). Conversely, at a fixed hot-zone count the ancilla overhead as a fraction of the data register diminishes with scale and remains manageable. The overall circuit volume showcases that four hot zones leads to advantageous performance of the OQFT at sizes around $512$ bits (Fig.~\ref{fig:OQFT_finals}(c)).

Finally, we address parallelism constraints. Because magic consumption and pipelining are stochastic, the number of simultaneously active qubits (DOF) varies over time. Figure~\ref{fig:OQFT_finals}(d) reports the average concurrent DOF at each system size. At sub-maximal hot-zone counts, the DOF increase over the regular QFT is again a constant multiplicative factor. Nearly regardless of system size, $\text{hz}=4$ requires \emph{on average} about $46$ concurrent operations. Aggregating the peak DOF requirement per hot zone for no-stalling Gidney adders, we find the total requirement of peak $128$ concurrent DOFs for $\text{hz}=4$. The gap between these peak and average DOF control values can be used as a design space for trading off control for time via stalling acceptance. For comparison purposes, the peak concurrent DOFs needs for the regular QFT would be of 32 DOF in our model.

Overall, we find that roughly $500$ logical ancillae and a $4\times$ increase in parallelism are needed to extract a substantial benefit (about $2\times$ lower runtime) when operating OQFTs relative to QFTs. These numbers persist across system sizes, defining parallelism increase as a meaningful hardware-integration target. Current atom shuttling technology relies on pairs of crossed acousto-optical deflectors (AODs)~\cite{Bluvstein_2023} which, from a bandwidth perspective, is compatible with transporting distance-$d=13$ surface-code patches required here~\cite{Manetsch_2025,Evered_2025}. Scaling to sustained $\sim128$ concurrent logical transport channels motivates continued co-design across atom-moving technology and QEC layouts to fully realize OQFT advantages; for example emerging technologies for large-scale atom control, once paired with coherent shuttling, could alleviate this challenge~\cite{Lin_2025}. Alternatively, one could relax the reaction-limited assumptions and trade peak throughput for simpler, but slower, implementations. That, however, may limit the application opportunities for OQFTs.
\section{Methods - Simulation model and validation}

Our quantitative results rest on two complementary models. At the \emph{micro} level, a tick-by-tick simulator advances one QEC cycle ($t_r = 1\;\text{ms}$) at a time, tracking the full state of a single adder invocation: qubit positions on a grid layout, bridge-qubit lifecycles (initialization, entanglement, consumption, reset), and magic-state availability. Factory output is stochastic, with parameters as described in Sec.~\ref{sec:OQFT_overview} leading to randomized T-state delivery across ticks. Every qubit in transit consumes one degree of freedom (DOF) per tick, with moves whose ballistic time exceed $t_r$ incurring an extra syndrome-extraction round. The simulator enforces a hard DOF cap per tick, and any operation that would exceed it is deferred. After the adder completes, total footprint, wall-clock time, per-tick DOF profile, and move-distance statistics are recorded. Multiple seeds per bit width yield mean and standard-deviation estimates.

At the \emph{macro} level, the OQFT evaluator composes full protocol schedules from a lookup table of per-bit-width adder statistics generated by the micro simulator. Given a system size $n$ and hot-zone count, the evaluator assigns hot zones to block pairs via a barrier scheduler, sums adder invocation times and inter-pair hop costs, and reports total time, footprint, and peak DOF for the complete OQFT or regular QFT. Because both adder variants are designed for stall-free, reaction-limited operation---factory patches are always either cultivating or holding a ready state, and data qubits remain stationary---the volume metric $V = Q_{\text{total}} \times t_{\text{total}}$ faithfully captures the space--time cost without idle-space discounting, providing a direct and micro-analyzed volume estimate in contrast with the high-level volume evaluation strategy used previously in Ref.~\cite{gidney2020quantumblocklookaheadadders}.

We validate the micro simulator through animated visualizations of individual adder runs, which confirm physically plausible routing, absence of deadlocks, and correct bridge/factory pipelining. Statistical consistency across independent seeds further confirms robustness. Validation animations and data necessary to reproduce our figures are available in Zenodo~\cite{oqft_data_zenodo_2026}.

\section{Conclusions and Outlook}

This work set out to answer a concrete co-design question: \emph{under which architectural requirements can the OQFT's depth-for-resources trade-off be made beneficial?} By jointly developing the algorithm's compilation and the underlying hardware model, from macro-level layout down to individual patch moves, we arrive at a clear answer for the surface-code, neutral-atom baseline studied here. The OQFT becomes beneficial once three architectural conditions are met: (i)~\emph{reaction-limited operation}, achieved through bridge-qubit interleaving and factory pipelining so that non-Clifford gates proceed at one per QEC tick; (ii)~\emph{resource mobility}, realized by mobile hot zones that bring magic, ancillae, and phase-gradient registers to stationary data; and (iii)~\emph{a considerable increased parallelism}, quantified at roughly $500$ additional logical ancillae and $128$ peak concurrent degrees of freedom to halve the execution time of a serial QFT. These requirements motivate further attention on control technology development by the community. Although these quantitative requirements are specific to the relatively resource-intensive OQFT, the underlying conditions above are more general. Any block-structured fault-tolerant routine using catalytic resource states faces the same three constraints, making the lessons broadly relevant to algorithm--architecture co-design. 

Several findings reinforce the value of micro-schedule-aware analysis. At the adder level, Gidney's and Cuccaro's ripple-carry constructions achieve comparable space--time volume once routing, bridge qubits, and factory footprint are jointly accounted for, contrasting with  high-level models which may underestimate these by $2$--$5\times$ at relevant register sizes. For us, the deciding factor is not volume but parallelism: Gidney's shallower MAJ and UMA blocks require substantially fewer concurrent degrees of freedom. At the protocol level, we observe a generic endianness mismatch challenge between the phase-gradient state and data registers. We resolve it here through cyclic PG swaps and alternating QFT reflections, techniques enabled by the reconfigurability of the neutral-atom platform. The five-layer OQFT structure then admits a clean trade-off: saturating the hot-zone count yields constant-time execution, while sub-maximal counts provide a predictable, near-linear speed-up per additional hot zone.

A recurring theme is the centrality of reaction-limited operation. Under transversal fault tolerance~\cite{Harry_AFT}, the QEC cycle coincides with the reaction time, and achieving this limit in practice demands coordination of magic delivery, bridge preparation, and qubit movement---details invisible at the circuit level but decisive for performance. Incorporating frame-update costs and gadgets such as Auto-T/CCZ~\cite{fowler2013timeoptimalquantumcomputation,gidney2019flexiblelayoutsurfacecode} more explicitly is a natural next step, moving toward the \emph{prepare--interact--resolve} synthesis advocated in Ref.~\cite{gidney2026magic} as a design principle for very-large-scale quantum integration.

Our analysis is scoped to cultivation-only magic-state factories and surface codes, limiting the directly applicable regime to OQFT registers up to ${\sim}2048$ bits for validation, and up to ${\sim}512$ bits for RSA factoring. Extending beyond would require distillation or lower physical error rates, each reshaping factory footprint and latency. Validating large OQFT instances is itself non-trivial: performance is conditioned on errors in the Frobenius metric, requiring sampling across the full Hilbert space; at our estimated ${\sim}10\text{s}$ per OQFT at $n=256$, direct validation is feasible only for modest sizes, and larger instances may be best tested inside a full multiplier pipeline or small factoring demonstrations.

More broadly, this work is intended as a case study in procedural routine--architecture co-design at a stage where systematic \emph{very-large-scale quantum systems integration} is still emerging. Just as early classical VLSI benefited from accumulating bespoke design exercises before crystallizing reusable abstractions~\cite{Lee1961AnAF,MeadConway1980VLSI,BrentKung1982ParallelAdders}, we believe the quantum community needs more such case studies where specific routines are tightly coupled with a concrete architectural model. This will enable the community to build the empirical base from which general design rules can eventually be extracted. The surface-code, neutral-atom setting we use is one well-understood baseline; repeating this exercise with high-rate qLDPC codes, alternative hardware platforms, and different fault-tolerance strategies is essential to identify which of the patterns we observe (reaction-limited bottlenecks, hot-zone mobility, parallelism as the adder differentiator) are fundamental and which are artifacts of the code or platform. Several concrete directions follow. Our hot-zone model can be refined with measurement-zone footprints~\cite{Bluvstein2025architecture}, continuous atom reloading~\cite{Chiu2025continuous}, and explicit syndrome-extraction budgets. Recent STAR architectures~\cite{ismail2025STAR} offer an alternative rotation-compilation strategy worth evaluating against the PG approach for early fault-tolerance demonstrations. On the adder front, our finding that reaction-limited operation narrows the volume gap between protocol families suggests that, with modestly higher magic-production rates, lookahead adders could become competitive---particularly attractive given their structured parallelism which can be tied qLDPC-efficient batching techniques available~\cite{xu2025batchedhighratelogicaloperations}. More broadly, the native reconfigurability of atom arrays is well suited to the non-local connectivity that qLDPC codes demand; realizing this potential requires systematizing the gadgets for magic generation and injection, and establishing code families whose architectural implications can be analyzed with the same rigor that surface codes currently afford. Finally, integrating the full OQFT into a complete multiplier evaluation, and asking whether the fixed five-layer structure can be exploited for batched transversal operations under high-rate codes, are natural next steps toward end-to-end performance assessment.

\section*{Acknowledgments}

We thank M. Kornjaca, C. Zhao, C. Duckering, A. Venkatramani, S. Ragavan, K. van Kirk, and G. Kahanamoku-Meyer for advice, suggestions, checking proposed opportunities to improve routing, information on experimental limits and helpful discussions. We thank S. Wang and S. Ostermann for creating an environment where this exploration was possible, and providing invaluable feedback on the early versions of this manuscript. We acknowledge the use of LLM tools to accelerate codebase development and refine narrative quality.

\section*{Data and code availability}

Validation animations and data accompanying this work are publicly archived on Zenodo~\cite{oqft_data_zenodo_2026}.

\bibliographystyle{IEEEtran}
\bibliography{references}

@misc{kahanamokumeyer2025logdepthinplacequantumfourier,
      title={A log-depth in-place quantum Fourier transform that rarely needs ancillas}, 
      author={Gregory D. Kahanamoku-Meyer and John Blue and Thiago Bergamaschi and Craig Gidney and Isaac L. Chuang},
      year={2025},
      eprint={2505.00701},
      archivePrefix={arXiv},
      primaryClass={quant-ph},
      url={https://arxiv.org/abs/2505.00701}, 
}

@misc{kahanamokumeyer2024fastquantumintegermultiplication,
      title={Fast quantum integer multiplication with zero ancillas}, 
      author={Gregory D. Kahanamoku-Meyer and Norman Y. Yao},
      year={2024},
      eprint={2403.18006},
      archivePrefix={arXiv},
      primaryClass={quant-ph},
      url={https://arxiv.org/abs/2403.18006}, 
}

@misc{fowler2013timeoptimalquantumcomputation,
      title={Time-optimal quantum computation}, 
      author={Austin G. Fowler},
      year={2013},
      eprint={1210.4626},
      archivePrefix={arXiv},
      primaryClass={quant-ph},
      url={https://arxiv.org/abs/1210.4626}, 
}

@misc{gidney2020quantumblocklookaheadadders,
      title={Quantum block lookahead adders and the wait for magic states}, 
      author={Craig Gidney},
      year={2020},
      eprint={2012.01624},
      archivePrefix={arXiv},
      primaryClass={quant-ph},
      url={https://arxiv.org/abs/2012.01624}, 
}

@article{Litinski2019magicstate,
  doi = {10.22331/q-2019-12-02-205},
  url = {https://doi.org/10.22331/q-2019-12-02-205},
  title = {Magic {S}tate {D}istillation: {N}ot as {C}ostly as {Y}ou {T}hink},
  author = {Litinski, Daniel},
  journal = {{Quantum}},
  issn = {2521-327X},
  publisher = {{Verein zur F{\"{o}}rderung des Open Access Publizierens in den Quantenwissenschaften}},
  volume = {3},
  pages = {205},
  month = dec,
  year = {2019}
}

@misc{litinski2022activevolumearchitectureefficient,
      title={Active volume: An architecture for efficient fault-tolerant quantum computers with limited non-local connections}, 
      author={Daniel Litinski and Naomi Nickerson},
      year={2022},
      eprint={2211.15465},
      archivePrefix={arXiv},
      primaryClass={quant-ph},
      url={https://arxiv.org/abs/2211.15465}, 
}

@misc{cuccaro2004,
      title={A new quantum ripple-carry addition circuit}, 
      author={Steven A. Cuccaro and Thomas G. Draper and Samuel A. Kutin and David Petrie Moulton},
      year={2004},
      eprint={quant-ph/0410184},
      archivePrefix={arXiv},
      primaryClass={quant-ph},
      url={https://arxiv.org/abs/quant-ph/0410184}, 
}

@misc{gidney2019windowedquantumarithmetic,
      title={Windowed quantum arithmetic}, 
      author={Craig Gidney},
      year={2019},
      eprint={1905.07682},
      archivePrefix={arXiv},
      primaryClass={quant-ph},
      url={https://arxiv.org/abs/1905.07682}, 
}

@misc{gidney2024magicstatecultivationgrowing,
      title={Magic state cultivation: growing T states as cheap as CNOT gates}, 
      author={Craig Gidney and Noah Shutty and Cody Jones},
      year={2024},
      eprint={2409.17595},
      archivePrefix={arXiv},
      primaryClass={quant-ph},
      url={https://arxiv.org/abs/2409.17595}, 
}

@misc{chen2025efficientmagicstatecultivation,
      title={Efficient Magic State Cultivation on $\mathbb{RP}^2$}, 
      author={Zi-Han Chen and Ming-Cheng Chen and Chao-Yang Lu and Jian-Wei Pan},
      year={2025},
      eprint={2503.18657},
      archivePrefix={arXiv},
      primaryClass={quant-ph},
      url={https://arxiv.org/abs/2503.18657}, 
}

@article{Gidney2018halvingcostof,
  doi = {10.22331/q-2018-06-18-74},
  url = {https://doi.org/10.22331/q-2018-06-18-74},
  title = {Halving the cost of quantum addition},
  author = {Gidney, Craig},
  journal = {{Quantum}},
  issn = {2521-327X},
  publisher = {{Verein zur F{\"{o}}rderung des Open Access Publizierens in den Quantenwissenschaften}},
  volume = {2},
  pages = {74},
  month = jun,
  year = {2018}
}

@misc{sahay2025foldtransversalsurfacecodecultivation,
      title={Fold-transversal surface code cultivation}, 
      author={Kaavya Sahay and Pei-Kai Tsai and Kathleen Chang and Qile Su and Thomas B. Smith and Shraddha Singh and Shruti Puri},
      year={2025},
      eprint={2509.05212},
      archivePrefix={arXiv},
      primaryClass={quant-ph},
      url={https://arxiv.org/abs/2509.05212}, 
}

@misc{menon2025magictricyclesefficientmagic,
      title={Magic tricycles: Efficient magic state generation with finite block-length quantum LDPC codes}, 
      author={Varun Menon and J. Pablo Bonilla-Ataides and Rohan Mehta and Andi Gu and Daniel Bochen Tan and Mikhail D. Lukin},
      year={2025},
      eprint={2508.10714},
      archivePrefix={arXiv},
      primaryClass={quant-ph},
      url={https://arxiv.org/abs/2508.10714}, 
}

@inproceedings{harry_architecture,
author = {Zhou, Hengyun and Duckering, Casey and Zhao, Chen and Bluvstein, Dolev and Cain, Madelyn and Kubica, Aleksander and Wang, Sheng-Tao and Lukin, Mikhail D.},
title = {Resource Analysis of Low-Overhead Transversal Architectures for Reconfigurable Atom Arrays},
year = {2025},
isbn = {9798400712616},
publisher = {Association for Computing Machinery},
address = {New York, NY, USA},
url = {https://doi.org/10.1145/3695053.3731039},
doi = {10.1145/3695053.3731039},
booktitle = {Proceedings of the 52nd Annual International Symposium on Computer Architecture},
pages = {1432–1448},
numpages = {17},
location = {
},
series = {ISCA '25}
}

@misc{gidney2025factor2048bitrsa,
      title={How to factor 2048 bit RSA integers with less than a million noisy qubits}, 
      author={Craig Gidney},
      year={2025},
      eprint={2505.15917},
      archivePrefix={arXiv},
      primaryClass={quant-ph},
      url={https://arxiv.org/abs/2505.15917}, 
}

@misc{Harry_AFT,
      title={Constant-Overhead Fault-Tolerant Quantum Computation with Reconfigurable Atom Arrays}, 
      author={Qian Xu and J. Pablo Bonilla Ataides and Christopher A. Pattison and Nithin Raveendran and Dolev Bluvstein and Jonathan Wurtz and Bane Vasic and Mikhail D. Lukin and Liang Jiang and Hengyun Zhou},
      year={2023},
      eprint={2308.08648},
      archivePrefix={arXiv},
      primaryClass={quant-ph},
      url={https://arxiv.org/abs/2308.08648}, 
}

@article{Lee1961AnAF,
  title={An Algorithm for Path Connections and Its Applications},
  author={C. Y. Lee},
  journal={IRE Trans. Electron. Comput.},
  year={1961},
  volume={10},
  pages={346-365},
  url={https://api.semanticscholar.org/CorpusID:40700386}
}

@article{BrentKung1982ParallelAdders,
  author  = {Brent, Richard P. and Kung, H. T.},
  title   = {A Regular Layout for Parallel Adders},
  journal = {IEEE Transactions on Computers},
  volume  = {C-31},
  number  = {3},
  pages   = {260--264},
  month   = mar,
  year    = {1982},
  doi     = {10.1109/TC.1982.1675982}
}

@book{MeadConway1980VLSI,
  author    = {Mead, Carver and Conway, Lynn},
  title     = {Introduction to {VLSI} Systems},
  publisher = {Addison-Wesley},
  address   = {Reading, MA},
  year      = {1980},
  isbn      = {0201043580}
}

@misc{gidney2019approximateencodedpermutationspiecewise,
      title={Approximate encoded permutations and piecewise quantum adders}, 
      author={Craig Gidney},
      year={2019},
      eprint={1905.08488},
      archivePrefix={arXiv},
      primaryClass={quant-ph},
      url={https://arxiv.org/abs/1905.08488}, 
}

@misc{qualtran,
      title={Expressing and Analyzing Quantum Algorithms with Qualtran}, 
      author={Matthew P. Harrigan and Tanuj Khattar and Charles Yuan and Anurudh Peduri and Noureldin Yosri and Fionn D. Malone and Ryan Babbush and Nicholas C. Rubin},
      year={2024},
      eprint={2409.04643},
      archivePrefix={arXiv},
      primaryClass={quant-ph},
      url={https://arxiv.org/abs/2409.04643}, 
}

@misc{coppersmith2002approximatefouriertransformuseful,
      title={An approximate Fourier transform useful in quantum factoring}, 
      author={D. Coppersmith},
      year={2002},
      eprint={quant-ph/0201067},
      archivePrefix={arXiv},
      primaryClass={quant-ph},
      url={https://arxiv.org/abs/quant-ph/0201067}, 
}

@misc{gidney2017approximateqftblog,
  author       = {Craig Gidney},
  title        = {Approximate Quantum Fourier Transform},
  year         = {2017},
  howpublished = {\url{https://algassert.com/post/1620}},
  note         = {Accessed: 2026-03-11},
}

@misc{qualtran2025approximateqft,
  author       = {{Quantumlib} and {Qualtran Contributors}},
  title        = {Approximate Quantum Fourier Transform},
  year         = {2025},
  url          = {https://github.com/quantumlib/Qualtran/blob/main/qualtran/bloqs/qft/approximate\_qft.py},
  note         = {Source code implementation in the Qualtran library},
}

@article{Gidney2021howtofactorbit,
  doi = {10.22331/q-2021-04-15-433},
  url = {https://doi.org/10.22331/q-2021-04-15-433},
  title = {How to factor 2048 bit {RSA} integers in 8 hours using 20 million noisy qubits},
  author = {Gidney, Craig and Eker{\aa{}}, Martin},
  journal = {{Quantum}},
  issn = {2521-327X},
  publisher = {{Verein zur F{\"{o}}rderung des Open Access Publizierens in den Quantenwissenschaften}},
  volume = {5},
  pages = {433},
  month = apr,
  year = {2021}
}

@misc{cain2025correlated,
      title={Correlated decoding of logical algorithms with transversal gates}, 
      author={Madelyn Cain and Chen Zhao and Hengyun Zhou and Nadine Meister and J. Pablo Bonilla Ataides and Arthur Jaffe and Dolev Bluvstein and Mikhail D. Lukin},
      year={2025},
      eprint={2403.03272},
      archivePrefix={arXiv},
      primaryClass={quant-ph},
      url={https://arxiv.org/abs/2403.03272}, 
}

@article{Anand_2024,
   title={A dual-species Rydberg array},
   volume={20},
   ISSN={1745-2481},
   url={http://dx.doi.org/10.1038/s41567-024-02638-2},
   DOI={10.1038/s41567-024-02638-2},
   number={11},
   journal={Nature Physics},
   publisher={Springer Science and Business Media LLC},
   author={Anand, Shraddha and Bradley, Conor E. and White, Ryan and Ramesh, Vikram and Singh, Kevin and Bernien, Hannes},
   year={2024},
   month=sep, pages={1744–1750} }

@misc{dutta2025,
      title={On Exact Space-Depth Trade-Offs in Multi-Controlled Toffoli Decomposition}, 
      author={Suman Dutta and Siyi Wang and Anubhab Baksi and Anupam Chattopadhyay and Subhamoy Maitra},
      year={2025},
      eprint={2502.01433},
      archivePrefix={arXiv},
      primaryClass={quant-ph},
      url={https://arxiv.org/abs/2502.01433}, 
}

@misc{gidney2019flexiblelayoutsurfacecode,
      title={Flexible layout of surface code computations using AutoCCZ states}, 
      author={Craig Gidney and Austin G. Fowler},
      year={2019},
      eprint={1905.08916},
      archivePrefix={arXiv},
      primaryClass={quant-ph},
      url={https://arxiv.org/abs/1905.08916}, 
}

@misc{litinski2023compute256bitellipticcurve,
      title={How to compute a 256-bit elliptic curve private key with only 50 million Toffoli gates}, 
      author={Daniel Litinski},
      year={2023},
      eprint={2306.08585},
      archivePrefix={arXiv},
      primaryClass={quant-ph},
      url={https://arxiv.org/abs/2306.08585}, 
}

@article{Gouzien_2023,
   title={Performance Analysis of a Repetition Cat Code Architecture: Computing 256-bit Elliptic Curve Logarithm in 9 Hours with 126133 Cat Qubits},
   volume={131},
   ISSN={1079-7114},
   url={http://dx.doi.org/10.1103/PhysRevLett.131.040602},
   DOI={10.1103/physrevlett.131.040602},
   number={4},
   journal={Physical Review Letters},
   publisher={American Physical Society (APS)},
   author={Gouzien, Elie and Ruiz, Diego and Le Regent, Francois-Marie and Guillaud, Jeremie and Sangouard, Nicolas},
   year={2023},
   month=jul }

@article{Bluvstein_2023,
   title={Logical quantum processor based on reconfigurable atom arrays},
   volume={626},
   ISSN={1476-4687},
   url={http://dx.doi.org/10.1038/s41586-023-06927-3},
   DOI={10.1038/s41586-023-06927-3},
   number={7997},
   journal={Nature},
   publisher={Springer Science and Business Media LLC},
   author={Bluvstein, Dolev and Evered, Simon J. and Geim, Alexandra A. and Li, Sophie H. and Zhou, Hengyun and Manovitz, Tom and Ebadi, Sepehr and Cain, Madelyn and Kalinowski, Marcin and Hangleiter, Dominik and Bonilla Ataides, J. Pablo and Maskara, Nishad and Cong, Iris and Gao, Xun and Sales Rodriguez, Pedro and Karolyshyn, Thomas and Semeghini, Giulia and Gullans, Michael J. and Greiner, Markus and Vuletić, Vladan and Lukin, Mikhail D.},
   year={2023},
   month=dec, pages={58–65} }

@misc{xu2025batchedhighratelogicaloperations,
      title={Batched high-rate logical operations for quantum LDPC codes}, 
      author={Qian Xu and Hengyun Zhou and Dolev Bluvstein and Madelyn Cain and Marcin Kalinowski and John Preskill and Mikhail D. Lukin and Nishad Maskara},
      year={2025},
      eprint={2510.06159},
      archivePrefix={arXiv},
      primaryClass={quant-ph},
      url={https://arxiv.org/abs/2510.06159}, 
}

@misc{xu2024fastparallelizablelogicalcomputation,
      title={Fast and Parallelizable Logical Computation with Homological Product Codes}, 
      author={Qian Xu and Hengyun Zhou and Guo Zheng and Dolev Bluvstein and J. Pablo Bonilla Ataides and Mikhail D. Lukin and Liang Jiang},
      year={2024},
      eprint={2407.18490},
      archivePrefix={arXiv},
      primaryClass={quant-ph},
      url={https://arxiv.org/abs/2407.18490}, 
}

@misc{yang2026spacetimeefficienthardwarecompatiblecomplexquantum,
      title={Spacetime-Efficient and Hardware-Compatible Complex Quantum Logic Units in qLDPC Codes}, 
      author={Willers Yang and Jason Chadwick and Mariesa H. Teo and Joshua Viszlai and Fred Chong},
      year={2026},
      eprint={2602.14273},
      archivePrefix={arXiv},
      primaryClass={quant-ph},
      url={https://arxiv.org/abs/2602.14273}, 
}

@misc{webster2026pinnaclearchitecturereducingcost,
      title={The Pinnacle Architecture: Reducing the cost of breaking RSA-2048 to 100 000 physical qubits using quantum LDPC codes}, 
      author={Paul Webster and Lucas Berent and Omprakash Chandra and Evan T. Hockings and Nouédyn Baspin and Felix Thomsen and Samuel C. Smith and Lawrence Z. Cohen},
      year={2026},
      eprint={2602.11457},
      archivePrefix={arXiv},
      primaryClass={quant-ph},
      url={https://arxiv.org/abs/2602.11457}, 
}

@article{Bluvstein2025architecture,
   title={A fault-tolerant neutral-atom architecture for universal quantum computation},
   volume={649},
   ISSN={1476-4687},
   url={http://dx.doi.org/10.1038/s41586-025-09848-5},
   DOI={10.1038/s41586-025-09848-5},
   number={8095},
   journal={Nature},
   publisher={Springer Science and Business Media LLC},
   author={Bluvstein, Dolev and Geim, Alexandra A. and Li, Sophie H. and Evered, Simon J. and Bonilla Ataides, J. Pablo and Baranes, Gefen and Gu, Andi and Manovitz, Tom and Xu, Muqing and Kalinowski, Marcin and Majidy, Shayan and Kokail, Christian and Maskara, Nishad and Trapp, Elias C. and Stewart, Luke M. and Hollerith, Simon and Zhou, Hengyun and Gullans, Michael J. and Yelin, Susanne F. and Greiner, Markus and Vuletić, Vladan and Cain, Madelyn and Lukin, Mikhail D.},
   year={2025},
   month=nov, pages={39–46} }

@misc{gidney2026magic,
  author       = {Gidney, Craig},
  title        = {How to eat magic states},
  howpublished = {Talk presented at the {APS} Global Summit 2026},
  year         = {2026},
  url          = {https://docs.google.com/presentation/d/1b0r3pKWi3_Bu64Rc5Ojc_9eVjWyZPWRP3-UBnqNdJB0},
  note         = {Slides available online}
}

@article{Chiu2025continuous,
   title={Continuous operation of a coherent 3,000-qubit system},
   volume={646},
   ISSN={1476-4687},
   url={http://dx.doi.org/10.1038/s41586-025-09596-6},
   DOI={10.1038/s41586-025-09596-6},
   number={8087},
   journal={Nature},
   publisher={Springer Science and Business Media LLC},
   author={Chiu, Neng-Chun and Trapp, Elias C. and Guo, Jinen and Abobeih, Mohamed H. and Stewart, Luke M. and Hollerith, Simon and Stroganov, Pavel L. and Kalinowski, Marcin and Geim, Alexandra A. and Evered, Simon J. and Li, Sophie H. and Lyu, Xingjian and Peters, Lisa M. and Bluvstein, Dolev and Wang, Tout T. and Greiner, Markus and Vuletić, Vladan and Lukin, Mikhail D.},
   year={2025},
   month=sep, pages={1075–1080} }

@misc{ismail2025STAR,
      title={Transversal STAR architecture for megaquop-scale quantum simulation with neutral atoms}, 
      author={Refaat Ismail and I-Chi Chen and Chen Zhao and Ronen Weiss and Fangli Liu and Hengyun Zhou and Sheng-Tao Wang and Andrew Sornborger and Milan Kornjača},
      year={2025},
      eprint={2509.18294},
      archivePrefix={arXiv},
      primaryClass={quant-ph},
      url={https://arxiv.org/abs/2509.18294}, 
}

@misc{cain2026shors10000,
      title={Shor's algorithm is possible with as few as 10,000 reconfigurable atomic qubits}, 
      author={Madelyn Cain and Qian Xu and Robbie King and Lewis R. B. Picard and Harry Levine and Manuel Endres and John Preskill and Hsin-Yuan Huang and Dolev Bluvstein},
      year={2026},
      eprint={2603.28627},
      archivePrefix={arXiv},
      primaryClass={quant-ph},
      url={https://arxiv.org/abs/2603.28627}, 
}

@misc{babbush2026ECC,
      title={Securing Elliptic Curve Cryptocurrencies against Quantum Vulnerabilities: Resource Estimates and Mitigations}, 
      author={Ryan Babbush and Adam Zalcman and Craig Gidney and Michael Broughton and Tanuj Khattar and Hartmut Neven and Thiago Bergamaschi and Justin Drake and Dan Boneh},
      year={2026},
      eprint={2603.28846},
      archivePrefix={arXiv},
      primaryClass={quant-ph},
      url={https://arxiv.org/abs/2603.28846}, 
}

@article{Jones4T_toffoli,
  title = {Low-overhead constructions for the fault-tolerant Toffoli gate},
  author = {Jones, Cody},
  journal = {Phys. Rev. A},
  volume = {87},
  issue = {2},
  pages = {022328},
  numpages = {4},
  year = {2013},
  month = {Feb},
  publisher = {American Physical Society},
  doi = {10.1103/PhysRevA.87.022328},
  url = {https://link.aps.org/doi/10.1103/PhysRevA.87.022328}
}

@dataset{oqft_data_zenodo_2026,
  author    = {Lopes, Pedro L. S.},
  title     = {{Data set, figure-generation script, and validation animations for
               "Deploying Optimistic Quantum Fourier Transforms: An Architecture-Algorithm Co-Design Study"}},
  year      = {2026},
  publisher = {Zenodo},
  doi       = {10.5281/zenodo.20028120},
  url       = {https://doi.org/10.5281/zenodo.20028120}
}

@article{Bluvstein_2022_coherent,
   title={A quantum processor based on coherent transport of entangled atom arrays},
   volume={604},
   ISSN={1476-4687},
   url={http://dx.doi.org/10.1038/s41586-022-04592-6},
   DOI={10.1038/s41586-022-04592-6},
   number={7906},
   journal={Nature},
   publisher={Springer Science and Business Media LLC},
   author={Bluvstein, Dolev and Levine, Harry and Semeghini, Giulia and Wang, Tout T. and Ebadi, Sepehr and Kalinowski, Marcin and Keesling, Alexander and Maskara, Nishad and Pichler, Hannes and Greiner, Markus and Vuletić, Vladan and Lukin, Mikhail D.},
   year={2022},
   month=Apr, pages={451–456} }

@misc{jones2013thesis,
      title={Logic Synthesis for Fault-Tolerant Quantum Computers}, 
      author={N. Cody Jones},
      year={2013},
      eprint={1310.7290},
      archivePrefix={arXiv},
      primaryClass={quant-ph},
      url={https://arxiv.org/abs/1310.7290}, 
}

@article{Takahashi:2005ygp,
    author = "Takahashi, Yasuhiro and Kunihiro, Noboru",
    title = "{A linear-size quantum circuit for addition with no ancillary qubits}",
    doi = "10.26421/QIC5.6-2",
    journal = "Quant. Inf. Comput.",
    volume = "5",
    number = "6",
    pages = "440--448",
    year = "2005"
}

@article{Evered_2025,
   title={Probing the Kitaev honeycomb model on a neutral-atom quantum computer},
   volume={645},
   ISSN={1476-4687},
   url={http://dx.doi.org/10.1038/s41586-025-09475-0},
   DOI={10.1038/s41586-025-09475-0},
   number={8080},
   journal={Nature},
   publisher={Springer Science and Business Media LLC},
   author={Evered, Simon J. and Kalinowski, Marcin and Geim, Alexandra A. and Manovitz, Tom and Bluvstein, Dolev and Li, Sophie H. and Maskara, Nishad and Zhou, Hengyun and Ebadi, Sepehr and Xu, Muqing and Campo, Joseph and Cain, Madelyn and Ostermann, Stefan and Yelin, Susanne F. and Sachdev, Subir and Greiner, Markus and Vuletić, Vladan and Lukin, Mikhail D.},
   year={2025},
   month=Sept, pages={341–347} }

@article{Manetsch_2025,
   title={A tweezer array with 6,100 highly coherent atomic qubits},
   volume={647},
   ISSN={1476-4687},
   url={http://dx.doi.org/10.1038/s41586-025-09641-4},
   DOI={10.1038/s41586-025-09641-4},
   number={8088},
   journal={Nature},
   publisher={Springer Science and Business Media LLC},
   author={Manetsch, Hannah J. and Nomura, Gyohei and Bataille, Elie and Lv, Xudong and Leung, Kon H. and Endres, Manuel},
   year={2025},
   month=Sept, pages={60–67} }

@article{Pause_2024,
   title={Supercharged two-dimensional tweezer array with more than 1000 atomic qubits},
   volume={11},
   ISSN={2334-2536},
   url={http://dx.doi.org/10.1364/OPTICA.513551},
   DOI={10.1364/optica.513551},
   number={2},
   journal={Optica},
   publisher={Optica Publishing Group},
   author={Pause, Lars and Sturm, Lukas and Mittenbühler, Marcel and Amann, Stephan and Preuschoff, Tilman and Schäffner, Dominik and Schlosser, Malte and Birkl, Gerhard},
   year={2024},
   month=Feb, pages={222} }

@article{fast_transport,
  title = {Fast neutral-atom transport and transfer between optical tweezers},
  author = {Cicali, Cristina and Calzavara, Martino and Cuestas, Eloisa and Calarco, Tommaso and Zeier, Robert and Motzoi, Felix},
  journal = {Phys. Rev. Appl.},
  volume = {24},
  issue = {2},
  pages = {024070},
  numpages = {23},
  year = {2025},
  month = {Aug},
  publisher = {American Physical Society},
  doi = {10.1103/7r3w-8m61},
  url = {https://link.aps.org/doi/10.1103/7r3w-8m61}
}

@article{Lin_2025,
   title={AI-Enabled Parallel Assembly of Thousands of Defect-Free Neutral Atom Arrays},
   volume={135},
   ISSN={1079-7114},
   url={http://dx.doi.org/10.1103/2ym8-vs82},
   DOI={10.1103/2ym8-vs82},
   number={6},
   journal={Physical Review Letters},
   publisher={American Physical Society (APS)},
   author={Lin, Rui and Zhong, Han-Sen and Li, You and Zhao, Zhang-Rui and Zheng, Le-Tian and Hu, Tai-Ran and Wu, Hong-Ming and Wu, Zhan and Ma, Wei-Jie and Gao, Yan and Zhu, Yi-Kang and Su, Zhao-Feng and Ouyang, Wan-Li and Zhang, Yu-Chen and Rui, Jun and Chen, Ming-Cheng and Lu, Chao-Yang and Pan, Jian-Wei},
   year={2025},
   month=Aug }

\end{document}